\documentclass[aps,pra,twocolumn,showpacs,floatfix,preprintnumbers,superscriptaddress,amsmath,amssymb]{revtex4-1}

\usepackage{graphicx}
\usepackage{dcolumn}% Align table columns on decimal point
\usepackage{units}
\usepackage[colorlinks]{hyperref}
\usepackage{natbib}

  \newcommand{\co}{(Color online)} % APS journals. Info about color
\newcommand{\ket}[1]{\left| #1 \right>} % for Dirac bras
\renewcommand{\section}[1]{\paragraph*{#1.}}
\renewcommand{\subsection}[1]{\paragraph*{#1.}}
\newcommand{\paren}[1]{\ensuremath{\left(#1\right)}}

\begin{document}

\title{Tunneling Theory for Tunable Open Quantum Systems of Ultracold Atoms in
  One-Dimensional Traps}

\author{R.~Lundmark} \affiliation{Department of Fundamental Physics,
  Chalmers University of Technology, SE-412 96 G\"oteborg, Sweden}

\author{C.~Forss\'en} \email{christian.forssen@chalmers.se}
\affiliation{Department of Fundamental Physics,
  Chalmers University of Technology, SE-412 96 G\"oteborg, Sweden}
\affiliation{Department of Physics and Astronomy, University of
  Tennessee, Knoxville, TN 37996, USA}
\affiliation{Physics Division, Oak Ridge National Laboratory, Oak Ridge,
  TN 37831, USA} 

\author{J.~Rotureau}
\affiliation{Department of Fundamental Physics,
  Chalmers University of Technology, SE-412 96 G\"oteborg, Sweden}

\begin{abstract} 
  The creation of tunable open quantum systems is becoming feasible in
  current experiments with ultracold atoms in low-dimensional
  traps. In particular, the high degree of experimental control over
  these systems allows detailed studies of tunneling dynamics, e.g.,
  as a function of the trapping geometry and the interparticle
  interaction strength. In order to address this exciting opportunity
  we present a theoretical framework for two-body tunneling based on
  the rigged Hilbert space formulation. In this approach, bound,
  resonant and scattering states are included on an equal footing, and
  we argue that the coupling of all these components is vital for a
  correct description of the relevant threshold phenomena. In
  particular, we study the tunneling mechanism for two-body systems in
  one-dimensional traps and different interaction regimes. We find a
  strong dominance of sequential tunneling of single particles for
  repulsive and weakly attractive systems, while there is a signature
  of correlated pair tunneling in the calculated many-particle flux
  for strongly attractive interparticle interaction.
\end{abstract}


\maketitle
%------------------------------------------------
\section{Introduction}\label{sec:intro}
%------------------------------------------------
%
The tunneling of particles, energetically confined by a potential
barrier, is a fascinating quantum phenomenon which plays an important
role in many physical systems.  In nuclear physics, the tunneling
process was first discussed in the context of alpha
decay~\cite{1928ZPhy...51..204G}. For multiparticle decay, the
emission process gets more involved as the interaction between the
emitted particles can strongly impact the decay probability.
The relative importance of sequential (i.e., successive
single-particle) and non-sequential decay channels is a pivotal
question for such many-body systems, and the general phenomenon of
pairing in fermionic systems~\cite{degennes1999, Bardeen:1957hw,
  Migdal:1959kl,bohr1998, 1969Natur.224..673B} becomes very relevant.
For example, the nuclear pairing interaction is known to enhance the
probability of two-proton radioactivity~\cite{Pfutzner:2012ig,
  Grigorenko:2000cd, Grigorenko:2007ja, Maruyama:2012hi}.
Furthermore, the Coulomb interaction between electrons plays a crucial
role in the double ionization of atoms~\cite{Pfeiffer:2011if,
  Walker:1994hb}, although a full theoretical understanding of this
two-body decay is still lacking.

An exciting recent development in the context of multiparticle
tunneling is the experimental realization of few-body Fermi systems
with ultracold atoms~\cite{Serwane:2011hp,2013PhRvL.111q5302Z}. These
setups are extremely versatile as they are associated with a high
degree of experimental control over key parameters such as the number
of particles and the shape of the confining potential. In addition,
the interaction between particles can be tuned using Feshbach
resonances~\cite{Chin:2010kl}, which in the case of trapped particles
turns into a confinement-induced
resonance~\cite{Olshanii:1998jr,Zuern:2013cr}. The resulting
interparticle interaction is of very short range compared to the size
of the systems, and can be modeled with high accuracy by a zero-range
potential.
Such tunable open quantum systems provides a unique opportunity to
investigate the mechanism of tunneling as a function of the trap
geometry and the strength of the interparticle interaction.

The dynamics of quantum tunneling can be naturally modeled
using time-dependent theoretical approaches. See for example
Refs.~\cite{Lode:2009eh,Kim:2011hq,Lode:2012be,Hunn:2013fq} for
studies of two-atom tunneling with repulsive interactions and idealized
geometries, and Refs.~\cite{Taniguchi:2011jl,Maruyama:2012hi} for attractive
interactions.
Different dynamical regimes of multiparticle tunneling through a thin
barrier was discussed in
Refs.~\cite{delCampo:2006wb,delCampo:2011jx,Pons:2012br} with a focus
on the  effects of quantum statistics.
In addition, the single-particle and pair tunneling of trapped
fermionic atoms with attractive interactions were recently studied
employing a time-independent quasiparticle
formalism~\cite{2012PhRvL.108k5302R,Rontani:2013dg} in which
the tunneling rate was obtained using the semiclassical
Wentzel-Kramers-Brillouin (WKB) approximation.
However, that approach suffers from the uncontrolled approximation of
artificially dividing the space into different regions. The
time-dependent approaches, on the other hand, are not very reliable
when the decay width is small and they are not easily extended to
many-particle systems.

The purpose of this Rapid Communication is to introduce an alternative approach to
the study of open quantum systems with ultracold atoms. Our method is
based on the rigged Hilbert space formulation, that extends beyond the domain
of hermitian quantum mechanics and includes also time-asymmetric processes such as
decays (see e.g.\ Ref.~\cite{delaMadrid:2005gt} and references
therein).
In nuclear physics this formulation has been
employed in the Gamow Shell Model~\cite{Michel:2002hk,
  Rotureau:2006fo, Hagen:2006fy,Michel:2009jg,Papadimitriou:2011ik,
  2013FBS....54..725R} to study threshold states and decay
processes. 
Recently, it has also been used to model near-threshold, bound states
of dipolar molecules~\cite{Fossez:2013fj}.
Here, we focus on the basic example of two interacting atoms
in a one-dimensional (1D) trap, and we present results that highlight the
importance of a unified treatment of bound, resonant and scattering
states for the proper description of tunneling phenomena in
ultracold atoms. We study in particular the decay mechanism, and
we perform realistic calculations to make comparisons with recent
experimental data~\cite{2013PhRvL.111q5302Z}.
%
%------------------------------------------------
\section{Theoretical Formalism}\label{sec:theory}
\begin{figure}
\includegraphics[width=\columnwidth]{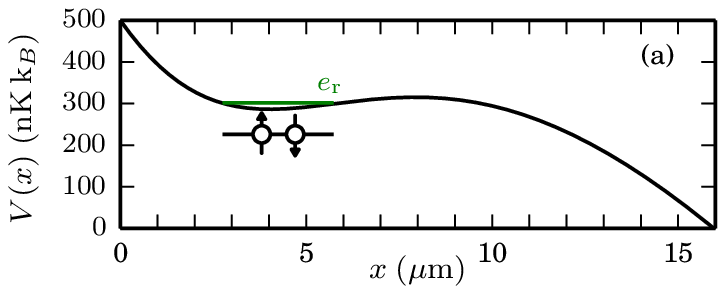}\\[-2ex]
\includegraphics[width=\columnwidth]{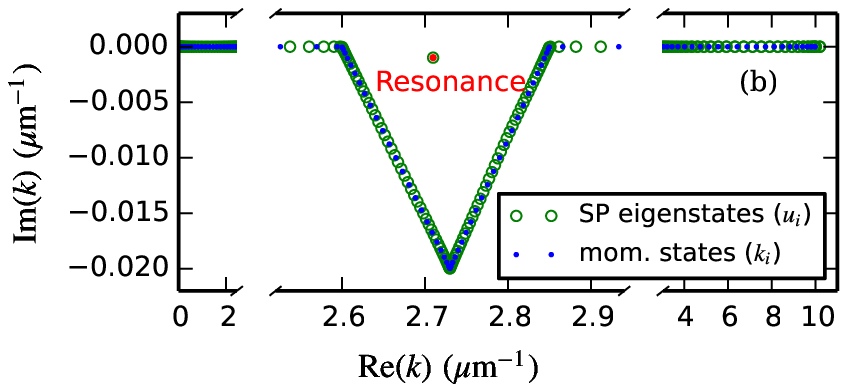}
%\resizebox{1.0\columnwidth}{!}{\input{fig/OneDContour}}
\caption{\label{fig:OneDContour}\co\ Panel (a): Trap potential, indicating
  the position of SP and two-body resonance states. Panel (b):
  Complex-momentum contour and Berggren basis
  states,  highlighting the position of the SP resonance pole.}
\end{figure}

%------------------------------------------------
We consider a system of interacting, two-component fermions in a
finite-depth potential trap.  The trap does not support any
single-particle (SP) bound states, but is just deep enough to support
a SP quasi-bound state with a finite lifetime. For
definiteness, we employ a 1D potential corresponding to the
experimental setup in Ref.~\cite{2013PhRvL.111q5302Z}, as illustrated in
Fig.~\ref{fig:OneDContour}(a). 
Let us denote this potential $V(x)$, with $x$ the degree of freedom in
the direction of the trap.  The interaction between fermionic
atoms in different hyperfine states is modeled by the zero-range
potential $V^{\delta}(x,x')=g \delta\paren{x - x'}$, with $g$ the
tunable interaction strength.  The fermions will be referred according
to their hyperfine spin state as ``spin-up'' ($\uparrow$) and
``spin-down'' ($\downarrow$), thus making an obvious connection with
systems of spin 1/2 particles (e.g. electrons or nucleons).
In this Rapid Communication we will restrict ourselves to the simplest instance
of such a tunable open quantum system, the case of two interacting
fermions in different spin states in an open 1D potential
trap. However, we want to stress that the formalism can be applied to
higher-dimensional traps and to systems with more particles.

The Hamiltonian for the two-particle system is
\begin{align}
H = \sum_{i=1}^{2} \left [ -\frac{\hbar^2}{2m} \frac{d^2}{dx_i^2} +
  V(x_i)\right ]+g \delta(x_1-x_2), \label{H2} 
\end{align}
with $m$ the mass of the particle.
Let us first consider the situation of two non-interacting particles,
i.e., $g=0$.  In this case, the ground state of the system,
$|\Phi^{(0)}\rangle$, corresponds to the two distinguishable fermions
both occupying the resonant (quasi-bound) state $|u_\mathrm{res}
\rangle$ of the SP Hamiltonian: $h(x) = -\hbar^2 / (2m) d^2 / dx^2
+ V(x)$. In this configuration, both particles are localized in the
trap for a finite amount of time, before tunneling out through the
potential barrier.  The decaying SP state $|u_\mathrm{res}\rangle$
can be described as a Gamow state~\cite{1928ZPhy...51..204G}. Such a
state behaves asympotically as an outgoing wave with a complex-energy
$e=e_{r}-i \gamma_r / 2$. The imaginary part of the energy corresponds
to the decay width $\gamma_r$ and gives the half-life of the SP
state, $t_{1/2}=\ln{(2)} \hbar / \gamma_r$, and the SP tunneling
rate $\gamma_1 = \gamma_r / \hbar$.

We will obtain solutions of the two-body Hamiltonian~\eqref{H2}, for
finite values of the interaction strength $g$ using an expansion of
SP states in the so-called Berggren
basis~\cite{Berggren:1968zz}. This complex-momentum basis includes
S-matrix poles (bound and resonant states) as well as non-resonant
scattering states associated with the potential $V(x)$. The use of
this basis is key to our approach as it allows to consistently include
the continuum when finding eigensolutions of the open quantum system.
It constitutes a rigged Hilbert space and the corresponding completeness relation is a
generalization of the Newton completeness
relation~\cite{1960JMP.....1..319N} (defined only for real energy
states) and reads
\begin{align}
\sum_n |u_n\rangle \langle \tilde{u}_n|+\int_{L^+} dk | u_k\rangle \langle
\tilde{u}_k|=1, \label{eq:berg_comp} 
\end{align}
where $|u_n\rangle$ correspond to poles of the S-matrix, and the
integral of states along the contour $L^+$, extending below the
resonance poles in the fourth
quadrant of the complex-momentum plane, represents the contribution from the
non-resonant scattering continuum~\cite{Berggren:1968zz}.
In practice, the integral in~\eqref{eq:berg_comp} is discretized in
two steps. First, the contour $L^{+}$ is truncated at $k =
k_\mathrm{max}$ and each segment is spanned by a Gauss-Legendre mesh
that gives a finite set of complex-momentum states $\{ \ket{k_i}
\}$. In a second step, the SP Hamiltonian is diagonalized in order
to obtain a finite set of SP basis states $U_1 \equiv
\{|u_i\rangle\}$~\cite{Michel:2009jg}.  The two-particle basis $T_2$
is then naturally constructed from the SP basis for the spin-up and
-down fermions as $T_2 \equiv U_1(\uparrow) \otimes
U_1(\downarrow)$. For the SP states along the complex contour, the
wave function diverges for $x \rightarrow \infty$ and as a
consequence, the matrix elements of the two-body interaction in the
Berggren basis are not finite. We solve this issue by regularizing the
two-body matrix elements between states in $T_2$ using an expansion in
the harmonic oscillator (HO) basis~\cite{Hagen:2006fy}.
Note that our Hamiltonian~\eqref{H2} matrix in this rigged Hilbert
space will be non-Hermitian, but complex symmetric.  The spectrum will
include bound, resonant and scattering many-body states.
Resonance solutions, $|\Phi_\mathrm{res}\rangle$, are characterized by
outgoing boundary conditions and a complex energy $E=E_r-i \Gamma_r /
2$, where $\Gamma_r$ is the decay width due to the emission of particles
out of the trap.
The resonant solution will be identified in the many-body spectrum, as
the state with the largest overlap (in modulus) with
$|\Phi^{(0)}\rangle$, referred to as the pole
approximation~\cite{Michel:2002hk}.
With this goal in mind we employ the Davidson algorithm for
diagonalization~\cite{Davidson:1975db,Michel:606452} which allows to target a
desired eigenpair.
Note that results for low-energy resonances should be independent of
the particular choice of ${L^+}$ as long as the Berggren completeness
relation~\eqref{eq:berg_comp} holds, i.e., $k_\mathrm{max}$ and the
number of discretization points both need to be large enough.

Concerning the tunneling rate we want to stress that there is {\it a
  priori} no simple relation between the decay width and the half life
for a many-body system, contrary to the case of a SP Gamow
state. Assuming exponential decay we would estimate the tunneling rate
$\gamma_\Gamma = \Gamma_r / \hbar = -2\mathrm{Im}(E) / \hbar$.
However, having access to the resonant wave function,
$\Phi_\mathrm{res}(x_1,x_2) \equiv \Phi_\mathrm{res}(\mathbf{x})$, we
can alternatively compute the decay rate using an integral
formalism~\cite{Grigorenko:2007ja}. The rate of particle emissions can
be obtained by integrating the outward flux of particles at large
distance ${x_\mathrm{out}}$ from the center of the trap, and
normalizing by the number of particles on the inside
\begin{equation}
\begin{split}
\gamma_{\mathrm{flux}} = 
\frac{\hbar}{i m N({{x_\mathrm{out}}})}
\sum_i \int_0^{{x_\mathrm{out}}} \prod_{j \neq i} dx_j
\left[ \Phi_\mathrm{res}^{*}(\mathbf{x})
  \frac{d}{dx_i}\Phi_\mathrm{res}(\mathbf{x}) \right. \\ 
 \left. -\left( \frac{d}{dx_i}\Phi_\mathrm{res}^{*}(\mathbf{x}) \right)
  {\Phi_\mathrm{res}(\mathbf{x})} \right]_{x_i={{x_\mathrm{out}}}},  \label{eq:tunnel} 
\end{split}
\end{equation}
with $N({x_\mathrm{out}}) = \int_{0}^{x_\mathrm{out}} \prod_j dx_j
|\Phi_\mathrm{res}(\mathbf{x})|^2$.
%
%------------------------------------------------
\section{Results}\label{sec:results}
%------------------------------------------------
In the experimental setup of Ref.~\cite{2013PhRvL.111q5302Z}, the
fermions are trapped in an effective 1D optical trap created by a
tightly focused laser beam (Rayleigh range $x_R$) combined with a
linear magnetic potential (magnetic field gradient $B'$) giving the
potential
\begin{align}
V(x) = p V_0 \paren{1-\frac{1}{1 + \paren{\frac{x}{x_R}}^2 }}
-c_{B,\sigma}\mu_B B' x. \label{eq:Vexp}
\end{align}
The depth $p V_0$ of the trapping potential depends on the number of
particles that are in the trap. In addition, the parameter $c_{B
  ,\sigma} \approx 1$, although the exact value depends on both the
magnetic-field strength and the spin of the particle.
For comparison with experimental
results we will use molecular units, in which energy is given in
$\unit{nK \, k_B}$, time in $\unit{\mu s}$, and distances in
$\unit{\mu m}$. In these units we have $\hbar =  \unit[7638.2]{nK \, k_B \,
\mu s}$, the Bohr magneton $\mu_B = \unit[6.7171 \cdot 10^{8}]{nK \, k_B \,
T^{-1}}$  and $\hbar^2/m = \unit[80.645]{nK \, k_B \, \mu m^2}$, where $m$
is the mass of a ${}^6$Li atom. 

In Fig.~\ref{fig:OneDContour}(a) we show for illustrative purpose the
trap potential with $p V_{0}=\unit[2.123 \cdot 10^{3}]{nK \, k_B}$,
$x_R=\unit[9.975]{\mu m}$, $B'=\unit[18.90 \cdot 10^{-8}]{T \, \mu m^{-1}}$,
$c_{B ,\sigma}=1 $ which closely resemble the parameters extracted from
experimental data (see also discussion below).
In order to handle the linear term $B' x$ we truncate the potential at
$x_\mathrm{cut}$, sufficiently far away from the relevant trap
region. In practice, this is achieved by applying a positive energy
shift $E_\mathrm{shift}$ so that $V(x_\mathrm{cut}) + E_\mathrm{shift}
= 0$. The energy shift is subtracted at the end, and we have verified
that the fluctuations in the SP energy (tunneling rate) with the choice of
$E_\mathrm{shift}$ was less than $\unit[0.04]{\%}$ ($\unit[2]{\%}$).

The SP Schr\"odinger equation is solved using the method described
above. The discrete set of complex-momentum states $\{
\ket{k_i} \}$ that span the contour $L^{+}$ is shown as blue dots in
Fig.~\ref{fig:OneDContour}(b). The 
energy shift that was used is $E_\mathrm{shift} =  \unit[500]{nK \,
  k_B}$.
The resulting set of eigenstates (green circles) lies very close to
the contour with the exception of one isolated state. The former
states correspond to non-resonant scattering solutions, while the
latter is a resonance.  Together, these eigenstates form the complete
set of SP basis states, $\{ \ket{u_i} \}$, that will be used in the
many-body calculation.

The number of points on the contour is increased until convergence of
the SP resonance energy is achieved.
Note that the resonance pole will
always remain fixed while the set of scattering states will depend on
the choice of the contour $L^{+}$. For illustration purposes, the
contour shown in Fig.~\ref{fig:OneDContour} consists of only
$N_\mathrm{pts}=100$ basis states while full calculations were
performed with $N_\mathrm{pts} = \text{240--320}$.
For this set of potential parameters we find $e = (301.415 -0.085 i)
\; \unit{nK \, k_B}$, which translates into a tunneling rate $\gamma_1 =
22.38 \; s^{-1}$.

\begin{figure}
\includegraphics[width=\columnwidth]{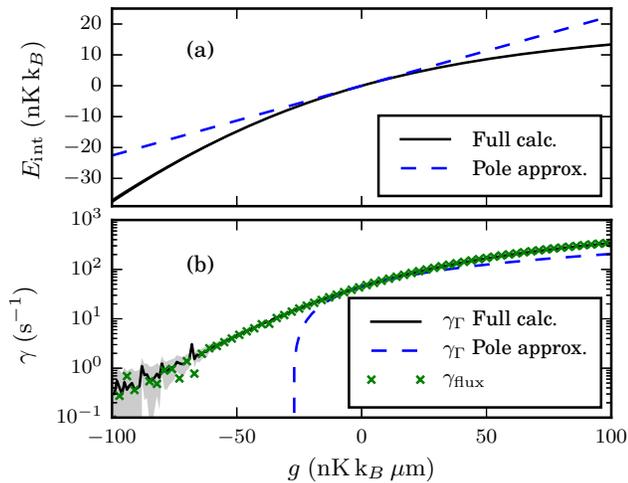}
\caption{\label{fig:ReERateStrength}\co\ Two-fermion resonance state as a
  function of the interaction strength $g$ for $c_B = 1$. Panel (a):
  Interaction energy~\eqref{eq:Eint} compared with the corresponding
  energy obtained using the pole approximation. Panel
  (b): Tunneling rates obtained from the imaginary part of the
  resonance energy (from the full calculation and the pole
  approximation, respectively) compared with the rate obtained from
  the flux calculation~\eqref{eq:tunnel}. }
\end{figure}

We now consider the solution of the interacting two-fermion system,
projected on the full Berggren basis. We define the interaction energy as
\begin{equation}
E_\mathrm{int} \equiv \mathrm{Re}(E) - 2 e_r ,
\label{eq:Eint}
\end{equation}
where $\mathrm{Re}(E) = E_r$ is the real part of the resonance energy.
Results for the two-particle resonance state as a function of the
interaction strength $g$ are shown in Fig.~\ref{fig:ReERateStrength}.
For $g=0$, the two fermions tunnel out independently and the tunneling
rate is equal to $\gamma = 2 \gamma_1 = 44.76~s^{-1}$. However, as the
interaction becomes more attractive, the real part of the resonance
energy decreases, and the effective barrier seen by the
two particles increases. As a consequence the tunneling rate
decreases as seen in Fig.~\ref{fig:ReERateStrength}(b).

Along with the full calculations, we show in
Fig.~\ref{fig:ReERateStrength} also results obtained in the pole
approximation, which corresponds to the single configuration where the
two distinguishable fermions occupy the SP resonant state. This
comparison clearly demonstrates the importance of continuum
correlations. The resonance energy and width are both decreased due to
configuration mixing between the SP resonance pole and non-resonant
scattering states. In particular, the energy width, which translates
into a decay rate, is very sensitive to these correlations. These
results highlight the importance of properly taking the openness of
the system into account.

The agreement between the tunneling rate computed from the decay width
of the resonance and from the flux formula~\eqref{eq:tunnel}
demonstrates the quality of our numerical approach. It also shows that
the tunneling is well approximated by an exponential decay law for
this system.
The numerical precision of results obtained in our approach was
studied in a series of convergence studies for systems with different
interaction strengths ($g = \unit[+100, -20, -100]{nK \, k_B \, \mu
  m}$). We varied the number of discretization points, modified the
contour in the complex-momentum plane, or changed the number of HO
states in the computation of interaction matrix elements(for a
detailed discussion of these studies, see
Ref.~\cite{forssen2015proc}). 
An uncertainty on the numerical results for a specific coupling
coefficient was then extracted based on the amplitudes of variations
when these model-space parameters were varied one by one. Adding these
amplitudes in quadrature gave an uncertainty in the real part of the
interaction energy on the order of $\lesssim \unit[2]{\%}$ for
the entire range of interaction strengths. 
However, the precision of the computed imaginary energy was found to have a
lower bound since variations of the computed decay rate was never
smaller than \unit[0.5]{$s^{-1}$}. This becomes obvious when the
interaction is strongly attractive and the ratio of imaginary and real
energies turns out to be very small. On the other hand, for larger
decay rates the variations were on the order of $\lesssim
\unit[1]{\%}$. In combination we have $\Delta \gamma = \max(0.01\gamma,
\unit[0.5]{s^{-1}})$.
This observation can be qualitatively understood in the following way:
The precision of the computed (complex) energy is not strongly
dependent on the value of $g$. For the most repulsive interactions,
the values of the real and imaginary parts are much larger than this
precision. However, as the interaction becomes more attractive, the
imaginary part rapidly decreases. With the precision of the total
(complex) energy almost constant, this creates a much larger
(relative) uncertainty for the tunneling rate in this region. The
estimated uncertainties from these numerical studies are shown as
shaded bands in both panels of Fig.~\ref{fig:ReERateStrength}, but is
only visible in the tunneling rate for the most attractive
interactions ($g \lesssim \unit[-60]{nK \, k_B \, \mu m} $).
%
%------------------------------------------------
\subsection{Density and Tunneling Mechanism}
\begin{figure}[!hptb]
\includegraphics[width=\columnwidth]{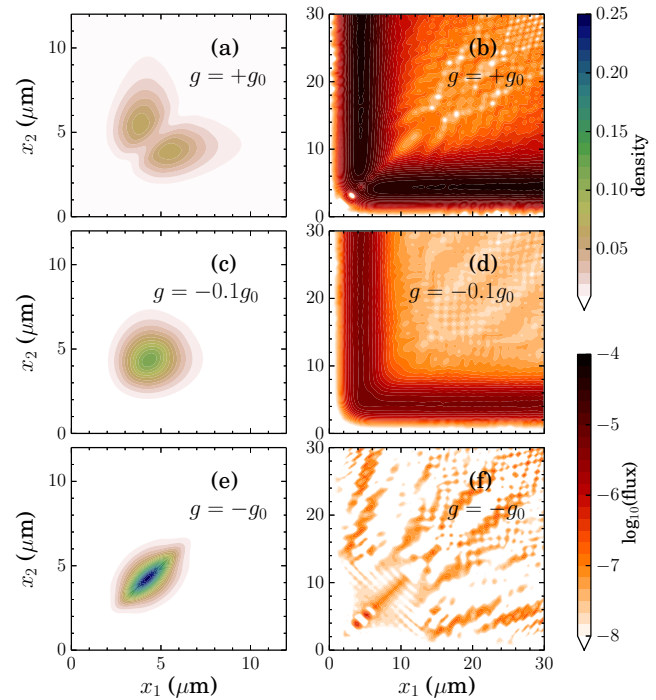}
\caption[Relative Difference]{\label{fig:DensFlux}\co\ Density contour
  plots (left panels) and logarithm of the outward particle flux
  (right panels) for repulsive ($g=+g_0$), slightly attractive ($g =
  -0.1 g_0$), and strongly attractive ($g = -g_0$) interactions (with
  $g_0 =  \unit[100]{nK \, k_B \, \mu m}$) from top to bottom,
  respectively.}
\end{figure}

%------------------------------------------------
%
The density and stationary outgoing particle flux can be seen in
Fig.~\ref{fig:DensFlux} for attractive and repulsive interactions. The
particles are localized around the trap minimum (at approximately $x =
\unit[4]{\mu m}$). For the repulsive interaction shown in 
Fig.~\ref{fig:DensFlux}(a) we can clearly observe the emerging
fermionization~\cite{Girardeau:1960ff,Gharashi:2013dg,Lindgren:2014kl}
of the two distinguishable particles by the development of an
$x_1=x_2$ valley in the density distribution.

The flux provides interesting insights into the tunneling
mechanism. For the repulsive and the slightly attractive scenarios,
shown in Figs.~\ref{fig:DensFlux}(b,d), the outgoing flux is mainly
concentrated in two bands, corresponding to one particle staying in
the trap and the other one leaving it. This indicates a strong
predominance of sequential tunneling. However, for the most attractive
case, shown in Fig.~\ref{fig:DensFlux}(f), we have significant outward
flux in the $x_1 \approx x_2$ region. This signals that the two
fermions can leave the trap simultaneously at short distance from each
other. Unfortunately, the region of strong attraction, where pair
tunneling appears as an important decay channel, is also characterized
by the smallest total tunneling rates. Therefore, there is significant
numerical noise for these particular results. We stress, however, that
the general conclusion of increasing pair tunneling remains true
although quantitative results cannot be obtained.
%
%------------------------------------------------
\subsection{Comparison with Experimental Data%
\label{sec:exp}}
%------------------------------------------------
%
The tunneling of few fermions from low-dimensional traps were measured
by \textcite{2013PhRvL.111q5302Z}. 
The experimental trap was highly elongated with a much
stronger confinement in the perpendicular direction,
$\omega_\parallel/\omega_\perp \approx 1/10$. Still, for very strong
attraction the size of the dimer can become comparable to
$b_\perp$, the length scale of the ignored (transverse) trap
dimensions.  In such a situation, the 1D approximation could
be questioned. Following Ref.~\cite{Gharashi:2013dg} we have verified
that we remain in the effectively one-dimensional region with $a_{1D}
/ b_\perp \approx 3$ for $g = \unit[-100]{nK \, k_B \, \mu m}$,
which is the largest attraction considered in this Rapid Communication.

The data analysis of Ref.~\cite{2013PhRvL.111q5302Z} is quite
complicated and involves the use of the WKB approximation to extract
the trap potential parameters. More precisely, $p V_0$ and $B'$ in
Eq.~\eqref{eq:Vexp} were adjusted such that the SP tunneling rates
obtained in the WKB approximation matched the experimental results.
Using the set of parameters given in Ref.~\cite{2013PhRvL.111q5302Z}
as input to our exact diagonalization approach leads to good agreement
for the SP energies (with a difference of at most a few percent),
while SP tunneling rates were almost two times larger than the
ones published in Ref.~\cite{2013PhRvL.111q5302Z}.
As a consequence, we have adopted the strategy of refitting the
parameters $p$ and $B'$ to reproduce measured SP tunneling rates.
Resulting changes of these parameters, compared to the WKB analysis, is
in the order of $\sim \unit[0.1]{\%}$.
From this one can conclude that the tunneling rate is very sensitive
to small shifts in the trapping potential, that continuum couplings
are very important, and that the uncontrolled WKB approximation may be
inadequate to use in a fitting procedure.

Using this new set of parameters, we compute the energies and the
tunneling rates for the two-particle system. For these calculations we
used a complex-momentum contour with slightly fewer discretization
points ($N_\mathrm{pts}=200$). Our predictions are
presented in Table~\ref{tab:TwoParticleRes} compared to experimental
data. 
\begin{table}[!htbp]
  \begin{ruledtabular}
  \begin{tabular}{cccc}
    $g$ $(\unit{nK \, k_B \, \mu m})$ & 
      $E_\mathrm{int}$ $(\unit{nK \, k_B})$ &
      $\gamma_{\Gamma}$ $(\unit{s^{-1}})$ &
      $\gamma_\mathrm{Exp}$ $(\unit{s^{-1}})$ \\ \hline 
%g=-31
    %\num[round-precision=1]{-30.969328218922707} &
    %\num[round-precision=3]{-8.449477437704743} &
    %\num[round-precision=1]{19.192048901280312} &
    -31.0 &
    \phantom{0}-8.4(2) &
    19.2(5) &
    22.2(10)\\ 
%g=-41
    %\num[round-precision=1]{-41.5270537481009} &
    %\num[round-precision=3]{-12.1037} & 
    %\num[round-precision=1]{12.5427}  &
    -41.5 &
    -12.1(3) &
    12.5(5) &
    13.8(10)
    \\ 
%g=-45
    %\num[round-precision=1]{-45.0462955911603} &
    %\num[round-precision=3]{-13.585} &
    %\num[round-precision=1]{25.809155170397453} &
    -45.0 &
    -13.6(3) &
    25.8(5) &
    \phantom{0}9.7(3)
    \\
%g=-100
    %\num[round-precision=1]{-99.94646834288692} &
    %\num[round-precision=3]{-37.023977651353455} &
    %\phantom{0}\num[round-precision=1]{0.436552419187243} &
    -99.9 &
    -37.0(7) &
    \phantom{0}0.4(5) &
    \phantom{0}2.14(20)
  \end{tabular}
  \end{ruledtabular}
  \caption[Two-Particle Result]{\label{tab:TwoParticleRes}Energy and
    tunneling rate for two atoms in a trap as a function of the
    interaction strength $g$. Experimental results from
    \textcite{2013PhRvL.111q5302Z}. Each case corresponds to a
    different trapping potential, as described in the text.} 
\end{table} 
The calculated tunneling rates are in acceptable agreement with the
measured ones. However, for one case ($g= \unit[-45.0]{nK \, k_B \, \mu
m}$) the difference is almost a factor three, which is well beyond the
expected precision of our method as indicated by the uncertainty
estimates of the tabulated results.
The fact that our tunneling rate is not monotonically decreasing as
the interaction becomes more attractive is due to the extreme
dependence on the SP potential. In particular, for each value of
$g$, the parameter $c_{B,\sigma}$ is slightly different. Moreover, the
spin dependence of this term gives rise to slightly different trapping
potentials for the two atoms. Unfortunately, this parameter is not
determined uniquely by the SP tunneling data and we have therefore
used $c_{B,\sigma}(g)$ as published in
Ref.~\cite{2013PhRvL.111q5302Z}. A better agreement with the
experimental results can certainly be achieved by relaxing the
predictive ambitions and tuning this parameter for each specific
interaction strength.
As a final note, our calculated interaction energies are about three
times larger than the values extracted (using WKB) from experiment. We
conclude that the WKB method should not be expected to
produce reliable estimates for this quantity and that the analysis of
experimental results for open quantum systems is highly sensitive to
the determination of trap parameters.
%
%------------------------------------------------
\section{Conclusion}\label{sec:conclusion}
%------------------------------------------------
In this Rapid Communication we have introduced the rigged Hilbert space formalism to
the theoretical study of tunneling in systems of ultracold atoms. We
focused on the case of two distinguishable particles in a
one-dimensional trap.  The two-atom dynamics was solved for a wide
range of interaction strengths by using an expansion in the Berggren
basis. 
The computational cost of this approach is mainly associated with the
construction of the Hamiltonian matrix. Fortunately, the two-body
interaction matrix elements are directly proportional to the
interaction strength $g$ and will only have to be computed once for a
specific model-space truncation.
We computed the energy and lifetime of two-body resonant states, and
could highlight the importance of continuum correlations for the
proper description of such threshold phenomena. Moreover, we were able
to obtain the density and flux distributions. The analysis of the
outgoing particle flux indicated a predominance of sequential
single-particle tunneling, with signs of pair tunneling for strongly
attractive systems. The numerical robustness of our method was
discussed and our theoretical predictions were compared with
experimental results. We found a quantitative agreement for tunneling
rates in systems with attractive interactions. However, interaction
energies differed significantly from those extracted from experimental
data using a WKB approach~\cite{2013PhRvL.111q5302Z} and we emphasized
that these differences stem from the uncontrolled approximation
inherent to semiclassical approaches.  

Our approach offers a number of key features:
As we use an expansion in a SP basis, the particle
statistics of the many-body states is treated exactly and the 
method can be straightforwardly extended to systems with more atoms
and other shapes for the trapping potential. By working in a rigged
Hilbert space we actually compute the complex energies of true many-body
resonances, which gives us both the position and the width. The
numerical precision of the method will be limited by the relative
magnitude of real and imaginary energies. Still, the
two-body systems that are studied in this work range from strongly
repulsive to strongly attractive, and the associated decay rates span
three orders of magnitude.
%
%------------------------------------------------
\section{Acknowledgments}
  The research leading to these results has received funding from the
  European Research Council under the European Community's Seventh
  Framework Programme (FP7/2007-2013) / ERC grant agreement
  no.~240603, and the Swedish Foundation for International Cooperation
  in Research and Higher Education (STINT, IG2012-5158). The
  computations were performed on resources provided by the Swedish
  National Infrastructure for Computing (SNIC) at High-Performance
  Computing Center North (HPC2N) and at Chalmers Centre for
  Computational Science and Engineering (C3SE).  We thank the European
  Centre for Theoretical Studies in Nuclear physics and Related Areas (ECT*)
  in Trento, and the Institute for Nuclear Theory at the University of
  Washington, for their hospitality and partial support during the
  completion of this work. We are much indebted to D. Blume, 
  M. Zhukov, N. Zinner, and G. Z\"urn for stimulating discussions.

\bibliography{1dtunneling,1dtunneling-temp}

\begin{thebibliography}{43}
\expandafter\ifx\csname natexlab\endcsname\relax\def\natexlab#1{#1}\fi
\expandafter\ifx\csname bibnamefont\endcsname\relax
  \def\bibnamefont#1{#1}\fi
\expandafter\ifx\csname bibfnamefont\endcsname\relax
  \def\bibfnamefont#1{#1}\fi
\expandafter\ifx\csname citenamefont\endcsname\relax
  \def\citenamefont#1{#1}\fi
\expandafter\ifx\csname url\endcsname\relax
  \def\url#1{\texttt{#1}}\fi
\expandafter\ifx\csname urlprefix\endcsname\relax\def\urlprefix{URL }\fi
\providecommand{\bibinfo}[2]{#2}
\providecommand{\eprint}[2][]{\url{#2}}

\bibitem[{\citenamefont{Gamow}(1928)}]{1928ZPhy...51..204G}
\bibinfo{author}{\bibfnamefont{G.}~\bibnamefont{Gamow}},
  \bibinfo{journal}{Zeitschrift f{\"u}r Physik} \textbf{\bibinfo{volume}{51}},
  \bibinfo{pages}{204} (\bibinfo{year}{1928}).

\bibitem[{\citenamefont{De~Gennes}(1999)}]{degennes1999}
\bibinfo{author}{\bibfnamefont{P.}~\bibnamefont{De~Gennes}},
  \emph{\bibinfo{title}{Superconductivity Of Metals And Alloys}}, Advanced
  Books Classics Series (\bibinfo{publisher}{Westview Press},
  \bibinfo{year}{1999}), ISBN \bibinfo{isbn}{9780813345840}.

\bibitem[{\citenamefont{Bardeen et~al.}(1957)\citenamefont{Bardeen, Cooper, and
  Schrieffer}}]{Bardeen:1957hw}
\bibinfo{author}{\bibfnamefont{J.}~\bibnamefont{Bardeen}},
  \bibinfo{author}{\bibfnamefont{L.~N.} \bibnamefont{Cooper}},
  \bibnamefont{and} \bibinfo{author}{\bibfnamefont{J.~R.}
  \bibnamefont{Schrieffer}}, \bibinfo{journal}{Phys. Rev. Lett.}
  \textbf{\bibinfo{volume}{108}}, \bibinfo{pages}{1175} (\bibinfo{year}{1957}).

\bibitem[{\citenamefont{Migdal}(1959)}]{Migdal:1959kl}
\bibinfo{author}{\bibfnamefont{A.~B.} \bibnamefont{Migdal}},
  \bibinfo{journal}{Nucl. Phys.} \textbf{\bibinfo{volume}{13}},
  \bibinfo{pages}{655} (\bibinfo{year}{1959}).

\bibitem[{\citenamefont{Bohr and Mottelson}(1998)}]{bohr1998}
\bibinfo{author}{\bibfnamefont{A.}~\bibnamefont{Bohr}} \bibnamefont{and}
  \bibinfo{author}{\bibfnamefont{B.}~\bibnamefont{Mottelson}},
  \emph{\bibinfo{title}{Nuclear Structure}}, \bibinfo{number}{Vol. I and II}
  (\bibinfo{publisher}{World Scientific}, \bibinfo{year}{1998}), ISBN
  \bibinfo{isbn}{9789810239794}.

\bibitem[{\citenamefont{Baym et~al.}(1969)\citenamefont{Baym, Pethick, and
  Pines}}]{1969Natur.224..673B}
\bibinfo{author}{\bibfnamefont{G.}~\bibnamefont{Baym}},
  \bibinfo{author}{\bibfnamefont{C.}~\bibnamefont{Pethick}}, \bibnamefont{and}
  \bibinfo{author}{\bibfnamefont{D.}~\bibnamefont{Pines}},
  \bibinfo{journal}{Nature} \textbf{\bibinfo{volume}{224}},
  \bibinfo{pages}{673} (\bibinfo{year}{1969}).

\bibitem[{\citenamefont{Pf{\"u}tzner et~al.}(2012)\citenamefont{Pf{\"u}tzner,
  Karny, Grigorenko, and Riisager}}]{Pfutzner:2012ig}
\bibinfo{author}{\bibfnamefont{M.}~\bibnamefont{Pf{\"u}tzner}},
  \bibinfo{author}{\bibfnamefont{M.}~\bibnamefont{Karny}},
  \bibinfo{author}{\bibfnamefont{L.~V.} \bibnamefont{Grigorenko}},
  \bibnamefont{and} \bibinfo{author}{\bibfnamefont{K.}~\bibnamefont{Riisager}},
  \bibinfo{journal}{Rev. Mod. Phys.} \textbf{\bibinfo{volume}{84}},
  \bibinfo{pages}{567} (\bibinfo{year}{2012}).

\bibitem[{\citenamefont{Grigorenko et~al.}(2000)\citenamefont{Grigorenko,
  Johnson, Mukha, Thompson, and Zhukov}}]{Grigorenko:2000cd}
\bibinfo{author}{\bibfnamefont{L.~V.} \bibnamefont{Grigorenko}},
  \bibinfo{author}{\bibfnamefont{R.~C.} \bibnamefont{Johnson}},
  \bibinfo{author}{\bibfnamefont{I.~G.} \bibnamefont{Mukha}},
  \bibinfo{author}{\bibfnamefont{I.~J.} \bibnamefont{Thompson}},
  \bibnamefont{and} \bibinfo{author}{\bibfnamefont{M.~V.}
  \bibnamefont{Zhukov}}, \bibinfo{journal}{Phys. Rev. Lett.}
  \textbf{\bibinfo{volume}{85}}, \bibinfo{pages}{22} (\bibinfo{year}{2000}).

\bibitem[{\citenamefont{Grigorenko and Zhukov}(2007)}]{Grigorenko:2007ja}
\bibinfo{author}{\bibfnamefont{L.~V.} \bibnamefont{Grigorenko}}
  \bibnamefont{and} \bibinfo{author}{\bibfnamefont{M.~V.}
  \bibnamefont{Zhukov}}, \bibinfo{journal}{Phys. Rev. C}
  \textbf{\bibinfo{volume}{76}}, \bibinfo{pages}{014008}
  (\bibinfo{year}{2007}).

\bibitem[{\citenamefont{Maruyama et~al.}(2012)\citenamefont{Maruyama, Oishi,
  Hagino, and Sagawa}}]{Maruyama:2012hi}
\bibinfo{author}{\bibfnamefont{T.}~\bibnamefont{Maruyama}},
  \bibinfo{author}{\bibfnamefont{T.}~\bibnamefont{Oishi}},
  \bibinfo{author}{\bibfnamefont{K.}~\bibnamefont{Hagino}}, \bibnamefont{and}
  \bibinfo{author}{\bibfnamefont{H.}~\bibnamefont{Sagawa}},
  \bibinfo{journal}{Phys. Rev. C} \textbf{\bibinfo{volume}{86}},
  \bibinfo{pages}{044301} (\bibinfo{year}{2012}).

\bibitem[{\citenamefont{Pfeiffer et~al.}(2011)\citenamefont{Pfeiffer, Cirelli,
  Smolarski, D{\"o}rner, and Keller}}]{Pfeiffer:2011if}
\bibinfo{author}{\bibfnamefont{A.~N.} \bibnamefont{Pfeiffer}},
  \bibinfo{author}{\bibfnamefont{C.}~\bibnamefont{Cirelli}},
  \bibinfo{author}{\bibfnamefont{M.}~\bibnamefont{Smolarski}},
  \bibinfo{author}{\bibfnamefont{R.}~\bibnamefont{D{\"o}rner}},
  \bibnamefont{and} \bibinfo{author}{\bibfnamefont{U.}~\bibnamefont{Keller}},
  \bibinfo{journal}{Nat Phys} \textbf{\bibinfo{volume}{7}},
  \bibinfo{pages}{428} (\bibinfo{year}{2011}).

\bibitem[{\citenamefont{Walker et~al.}(1994)\citenamefont{Walker, Sheehy,
  DiMauro, Agostini, Schafer, and Kulander}}]{Walker:1994hb}
\bibinfo{author}{\bibfnamefont{B.}~\bibnamefont{Walker}},
  \bibinfo{author}{\bibfnamefont{B.}~\bibnamefont{Sheehy}},
  \bibinfo{author}{\bibfnamefont{L.~F.} \bibnamefont{DiMauro}},
  \bibinfo{author}{\bibfnamefont{P.}~\bibnamefont{Agostini}},
  \bibinfo{author}{\bibfnamefont{K.~J.} \bibnamefont{Schafer}},
  \bibnamefont{and} \bibinfo{author}{\bibfnamefont{K.~C.}
  \bibnamefont{Kulander}}, \bibinfo{journal}{Phys. Rev. Lett.}
  \textbf{\bibinfo{volume}{73}}, \bibinfo{pages}{1227} (\bibinfo{year}{1994}).

\bibitem[{\citenamefont{Serwane et~al.}(2011)\citenamefont{Serwane, Z{\"u}rn,
  Lompe, Ottenstein, Wenz, and Jochim}}]{Serwane:2011hp}
\bibinfo{author}{\bibfnamefont{F.}~\bibnamefont{Serwane}},
  \bibinfo{author}{\bibfnamefont{G.}~\bibnamefont{Z{\"u}rn}},
  \bibinfo{author}{\bibfnamefont{T.}~\bibnamefont{Lompe}},
  \bibinfo{author}{\bibfnamefont{T.~B.} \bibnamefont{Ottenstein}},
  \bibinfo{author}{\bibfnamefont{A.~N.} \bibnamefont{Wenz}}, \bibnamefont{and}
  \bibinfo{author}{\bibfnamefont{S.}~\bibnamefont{Jochim}},
  \bibinfo{journal}{Science} \textbf{\bibinfo{volume}{332}},
  \bibinfo{pages}{336} (\bibinfo{year}{2011}).

\bibitem[{\citenamefont{Z{\"u}rn
  et~al.}(2013{\natexlab{a}})\citenamefont{Z{\"u}rn, Wenz, Murmann,
  Bergschneider, Lompe, and Jochim}}]{2013PhRvL.111q5302Z}
\bibinfo{author}{\bibfnamefont{G.}~\bibnamefont{Z{\"u}rn}},
  \bibinfo{author}{\bibfnamefont{A.~N.} \bibnamefont{Wenz}},
  \bibinfo{author}{\bibfnamefont{S.}~\bibnamefont{Murmann}},
  \bibinfo{author}{\bibfnamefont{A.}~\bibnamefont{Bergschneider}},
  \bibinfo{author}{\bibfnamefont{T.}~\bibnamefont{Lompe}}, \bibnamefont{and}
  \bibinfo{author}{\bibfnamefont{S.}~\bibnamefont{Jochim}},
  \bibinfo{journal}{Phys. Rev. Lett.} \textbf{\bibinfo{volume}{111}},
  \bibinfo{pages}{175302} (\bibinfo{year}{2013}{\natexlab{a}}).

\bibitem[{\citenamefont{Chin et~al.}(2010)\citenamefont{Chin, Grimm, Julienne,
  and Tiesinga}}]{Chin:2010kl}
\bibinfo{author}{\bibfnamefont{C.}~\bibnamefont{Chin}},
  \bibinfo{author}{\bibfnamefont{R.}~\bibnamefont{Grimm}},
  \bibinfo{author}{\bibfnamefont{P.}~\bibnamefont{Julienne}}, \bibnamefont{and}
  \bibinfo{author}{\bibfnamefont{E.}~\bibnamefont{Tiesinga}},
  \bibinfo{journal}{Rev. Mod. Phys.} \textbf{\bibinfo{volume}{82}},
  \bibinfo{pages}{1225} (\bibinfo{year}{2010}).

\bibitem[{\citenamefont{Olshanii}(1998)}]{Olshanii:1998jr}
\bibinfo{author}{\bibfnamefont{M.}~\bibnamefont{Olshanii}},
  \bibinfo{journal}{Phys. Rev. Lett.} \textbf{\bibinfo{volume}{81}},
  \bibinfo{pages}{938} (\bibinfo{year}{1998}).

\bibitem[{\citenamefont{Z{\"u}rn
  et~al.}(2013{\natexlab{b}})\citenamefont{Z{\"u}rn, Lompe, Wenz, Jochim,
  Julienne, and Hutson}}]{Zuern:2013cr}
\bibinfo{author}{\bibfnamefont{G.}~\bibnamefont{Z{\"u}rn}},
  \bibinfo{author}{\bibfnamefont{T.}~\bibnamefont{Lompe}},
  \bibinfo{author}{\bibfnamefont{A.~N.} \bibnamefont{Wenz}},
  \bibinfo{author}{\bibfnamefont{S.}~\bibnamefont{Jochim}},
  \bibinfo{author}{\bibfnamefont{P.~S.} \bibnamefont{Julienne}},
  \bibnamefont{and} \bibinfo{author}{\bibfnamefont{J.~M.}
  \bibnamefont{Hutson}}, \bibinfo{journal}{Phys. Rev. Lett.}
  \textbf{\bibinfo{volume}{110}}, \bibinfo{pages}{135301}
  (\bibinfo{year}{2013}{\natexlab{b}}).

\bibitem[{\citenamefont{Lode et~al.}(2009)\citenamefont{Lode, Streltsov, Alon,
  Meyer, and Cederbaum}}]{Lode:2009eh}
\bibinfo{author}{\bibfnamefont{A.~U.~J.} \bibnamefont{Lode}},
  \bibinfo{author}{\bibfnamefont{A.~I.} \bibnamefont{Streltsov}},
  \bibinfo{author}{\bibfnamefont{O.~E.} \bibnamefont{Alon}},
  \bibinfo{author}{\bibfnamefont{H.-D.} \bibnamefont{Meyer}}, \bibnamefont{and}
  \bibinfo{author}{\bibfnamefont{L.~S.} \bibnamefont{Cederbaum}},
  \bibinfo{journal}{J. Phys. B} \textbf{\bibinfo{volume}{42}},
  \bibinfo{pages}{044018} (\bibinfo{year}{2009}).

\bibitem[{\citenamefont{Kim and Brand}(2011)}]{Kim:2011hq}
\bibinfo{author}{\bibfnamefont{S.}~\bibnamefont{Kim}} \bibnamefont{and}
  \bibinfo{author}{\bibfnamefont{J.}~\bibnamefont{Brand}}, \bibinfo{journal}{J.
  Phys. B} \textbf{\bibinfo{volume}{44}}, \bibinfo{pages}{195301}
  (\bibinfo{year}{2011}).

\bibitem[{\citenamefont{Lode et~al.}(2012)\citenamefont{Lode, Streltsov,
  Sakmann, Alon, and Cederbaum}}]{Lode:2012be}
\bibinfo{author}{\bibfnamefont{A.~U.~J.} \bibnamefont{Lode}},
  \bibinfo{author}{\bibfnamefont{A.~I.} \bibnamefont{Streltsov}},
  \bibinfo{author}{\bibfnamefont{K.}~\bibnamefont{Sakmann}},
  \bibinfo{author}{\bibfnamefont{O.~E.} \bibnamefont{Alon}}, \bibnamefont{and}
  \bibinfo{author}{\bibfnamefont{L.~S.} \bibnamefont{Cederbaum}},
  \bibinfo{journal}{PNAS} \textbf{\bibinfo{volume}{109}},
  \bibinfo{pages}{13521} (\bibinfo{year}{2012}).

\bibitem[{\citenamefont{Hunn et~al.}(2013)\citenamefont{Hunn, Zimmermann,
  Hiller, and Buchleitner}}]{Hunn:2013fq}
\bibinfo{author}{\bibfnamefont{S.}~\bibnamefont{Hunn}},
  \bibinfo{author}{\bibfnamefont{K.}~\bibnamefont{Zimmermann}},
  \bibinfo{author}{\bibfnamefont{M.}~\bibnamefont{Hiller}}, \bibnamefont{and}
  \bibinfo{author}{\bibfnamefont{A.}~\bibnamefont{Buchleitner}},
  \bibinfo{journal}{Phys. Rev. A} \textbf{\bibinfo{volume}{87}},
  \bibinfo{pages}{043626} (\bibinfo{year}{2013}).

\bibitem[{\citenamefont{Taniguchi and Sawada}(2011)}]{Taniguchi:2011jl}
\bibinfo{author}{\bibfnamefont{T.}~\bibnamefont{Taniguchi}} \bibnamefont{and}
  \bibinfo{author}{\bibfnamefont{S.~I.} \bibnamefont{Sawada}},
  \bibinfo{journal}{Phys. Rev. E} \textbf{\bibinfo{volume}{83}},
  \bibinfo{pages}{026208} (\bibinfo{year}{2011}).

\bibitem[{\citenamefont{del Campo et~al.}(2006)\citenamefont{del Campo,
  Delgado, Garc{\'\i}a-Calder{\'o}n, Muga, and Raizen}}]{delCampo:2006wb}
\bibinfo{author}{\bibfnamefont{A.}~\bibnamefont{del Campo}},
  \bibinfo{author}{\bibfnamefont{F.}~\bibnamefont{Delgado}},
  \bibinfo{author}{\bibfnamefont{G.}~\bibnamefont{Garc{\'\i}a-Calder{\'o}n}},
  \bibinfo{author}{\bibfnamefont{J.}~\bibnamefont{Muga}}, \bibnamefont{and}
  \bibinfo{author}{\bibfnamefont{M.}~\bibnamefont{Raizen}},
  \bibinfo{journal}{Phys. Rev. A} \textbf{\bibinfo{volume}{74}},
  \bibinfo{pages}{013605} (\bibinfo{year}{2006}).

\bibitem[{\citenamefont{del Campo}(2011)}]{delCampo:2011jx}
\bibinfo{author}{\bibfnamefont{A.}~\bibnamefont{del Campo}},
  \bibinfo{journal}{Phys. Rev. A} \textbf{\bibinfo{volume}{84}},
  \bibinfo{pages}{012113} (\bibinfo{year}{2011}).

\bibitem[{\citenamefont{Pons et~al.}(2012)\citenamefont{Pons, Sokolovski, and
  del Campo}}]{Pons:2012br}
\bibinfo{author}{\bibfnamefont{M.}~\bibnamefont{Pons}},
  \bibinfo{author}{\bibfnamefont{D.}~\bibnamefont{Sokolovski}},
  \bibnamefont{and} \bibinfo{author}{\bibfnamefont{A.}~\bibnamefont{del
  Campo}}, \bibinfo{journal}{Phys. Rev. A} \textbf{\bibinfo{volume}{85}},
  \bibinfo{pages}{022107} (\bibinfo{year}{2012}).

\bibitem[{\citenamefont{Rontani}(2012)}]{2012PhRvL.108k5302R}
\bibinfo{author}{\bibfnamefont{M.}~\bibnamefont{Rontani}},
  \bibinfo{journal}{Phys. Rev. Lett.} \textbf{\bibinfo{volume}{108}},
  \bibinfo{pages}{115302} (\bibinfo{year}{2012}).

\bibitem[{\citenamefont{Rontani}(2013)}]{Rontani:2013dg}
\bibinfo{author}{\bibfnamefont{M.}~\bibnamefont{Rontani}},
  \bibinfo{journal}{Phys. Rev. A} \textbf{\bibinfo{volume}{88}},
  \bibinfo{pages}{043633} (\bibinfo{year}{2013}).

\bibitem[{\citenamefont{de~la Madrid}(2005)}]{delaMadrid:2005gt}
\bibinfo{author}{\bibfnamefont{R.}~\bibnamefont{de~la Madrid}},
  \bibinfo{journal}{Eur. J. Phys.} \textbf{\bibinfo{volume}{26}},
  \bibinfo{pages}{287} (\bibinfo{year}{2005}).

\bibitem[{\citenamefont{Michel et~al.}(2002)\citenamefont{Michel, Nazarewicz,
  P{\l}oszajczak, and Bennaceur}}]{Michel:2002hk}
\bibinfo{author}{\bibfnamefont{N.}~\bibnamefont{Michel}},
  \bibinfo{author}{\bibfnamefont{W.}~\bibnamefont{Nazarewicz}},
  \bibinfo{author}{\bibfnamefont{M.}~\bibnamefont{P{\l}oszajczak}},
  \bibnamefont{and}
  \bibinfo{author}{\bibfnamefont{K.}~\bibnamefont{Bennaceur}},
  \bibinfo{journal}{Phys. Rev. Lett.} \textbf{\bibinfo{volume}{89}},
  \bibinfo{pages}{042502} (\bibinfo{year}{2002}).

\bibitem[{\citenamefont{Rotureau et~al.}(2006)\citenamefont{Rotureau, Michel,
  Nazarewicz, P{\l}oszajczak, and Dukelsky}}]{Rotureau:2006fo}
\bibinfo{author}{\bibfnamefont{J.}~\bibnamefont{Rotureau}},
  \bibinfo{author}{\bibfnamefont{N.}~\bibnamefont{Michel}},
  \bibinfo{author}{\bibfnamefont{W.}~\bibnamefont{Nazarewicz}},
  \bibinfo{author}{\bibfnamefont{M.}~\bibnamefont{P{\l}oszajczak}},
  \bibnamefont{and} \bibinfo{author}{\bibfnamefont{J.}~\bibnamefont{Dukelsky}},
  \bibinfo{journal}{Phys. Rev. Lett.} \textbf{\bibinfo{volume}{97}},
  \bibinfo{pages}{110603} (\bibinfo{year}{2006}).

\bibitem[{\citenamefont{Hagen et~al.}(2006)\citenamefont{Hagen, Hjorth-Jensen,
  and Michel}}]{Hagen:2006fy}
\bibinfo{author}{\bibfnamefont{G.}~\bibnamefont{Hagen}},
  \bibinfo{author}{\bibfnamefont{M.}~\bibnamefont{Hjorth-Jensen}},
  \bibnamefont{and} \bibinfo{author}{\bibfnamefont{N.}~\bibnamefont{Michel}},
  \bibinfo{journal}{Phys. Rev. C} \textbf{\bibinfo{volume}{73}},
  \bibinfo{pages}{064307} (\bibinfo{year}{2006}).

\bibitem[{\citenamefont{Michel et~al.}(2009)\citenamefont{Michel, Nazarewicz,
  P{\l}oszajczak, and Vertse}}]{Michel:2009jg}
\bibinfo{author}{\bibfnamefont{N.}~\bibnamefont{Michel}},
  \bibinfo{author}{\bibfnamefont{W.}~\bibnamefont{Nazarewicz}},
  \bibinfo{author}{\bibfnamefont{M.}~\bibnamefont{P{\l}oszajczak}},
  \bibnamefont{and} \bibinfo{author}{\bibfnamefont{T.}~\bibnamefont{Vertse}},
  \bibinfo{journal}{J. Phys. G} \textbf{\bibinfo{volume}{36}},
  \bibinfo{pages}{013101} (\bibinfo{year}{2009}).

\bibitem[{\citenamefont{Papadimitriou et~al.}(2011)\citenamefont{Papadimitriou,
  Kruppa, Michel, Nazarewicz, Ploszajczak, and
  Rotureau}}]{Papadimitriou:2011ik}
\bibinfo{author}{\bibfnamefont{G.}~\bibnamefont{Papadimitriou}},
  \bibinfo{author}{\bibfnamefont{A.~T.} \bibnamefont{Kruppa}},
  \bibinfo{author}{\bibfnamefont{N.}~\bibnamefont{Michel}},
  \bibinfo{author}{\bibfnamefont{W.}~\bibnamefont{Nazarewicz}},
  \bibinfo{author}{\bibfnamefont{M.}~\bibnamefont{Ploszajczak}},
  \bibnamefont{and} \bibinfo{author}{\bibfnamefont{J.}~\bibnamefont{Rotureau}},
  \bibinfo{journal}{Phys. Rev. C} \textbf{\bibinfo{volume}{84}},
  \bibinfo{pages}{051304} (\bibinfo{year}{2011}).

\bibitem[{\citenamefont{Rotureau and van Kolck}(2013)}]{2013FBS....54..725R}
\bibinfo{author}{\bibfnamefont{J.}~\bibnamefont{Rotureau}} \bibnamefont{and}
  \bibinfo{author}{\bibfnamefont{U.}~\bibnamefont{van Kolck}},
  \bibinfo{journal}{Few-Body Syst} \textbf{\bibinfo{volume}{54}},
  \bibinfo{pages}{725} (\bibinfo{year}{2013}).

\bibitem[{\citenamefont{Fossez et~al.}(2013)\citenamefont{Fossez, Michel,
  Nazarewicz, and P{\l}oszajczak}}]{Fossez:2013fj}
\bibinfo{author}{\bibfnamefont{K.}~\bibnamefont{Fossez}},
  \bibinfo{author}{\bibfnamefont{N.}~\bibnamefont{Michel}},
  \bibinfo{author}{\bibfnamefont{W.}~\bibnamefont{Nazarewicz}},
  \bibnamefont{and}
  \bibinfo{author}{\bibfnamefont{M.}~\bibnamefont{P{\l}oszajczak}},
  \bibinfo{journal}{Phys. Rev. A} \textbf{\bibinfo{volume}{87}},
  \bibinfo{pages}{042515} (\bibinfo{year}{2013}).

\bibitem[{\citenamefont{Berggren}(1968)}]{Berggren:1968zz}
\bibinfo{author}{\bibfnamefont{T.}~\bibnamefont{Berggren}},
  \bibinfo{journal}{Nucl. Phys.} \textbf{\bibinfo{volume}{A109}},
  \bibinfo{pages}{265} (\bibinfo{year}{1968}).

\bibitem[{\citenamefont{Newton}(1960)}]{1960JMP.....1..319N}
\bibinfo{author}{\bibfnamefont{R.~G.} \bibnamefont{Newton}},
  \bibinfo{journal}{J. Math. Phys.} \textbf{\bibinfo{volume}{1}},
  \bibinfo{pages}{319} (\bibinfo{year}{1960}).

\bibitem[{\citenamefont{Davidson}(1975)}]{Davidson:1975db}
\bibinfo{author}{\bibfnamefont{E.~R.} \bibnamefont{Davidson}},
  \bibinfo{journal}{J. Comput. Phys.} \textbf{\bibinfo{volume}{17}},
  \bibinfo{pages}{87} (\bibinfo{year}{1975}).

\bibitem[{\citenamefont{Michel et~al.}(2003)\citenamefont{Michel, Nazarewicz,
  Ploszajczak, and Okolowicz}}]{Michel:606452}
\bibinfo{author}{\bibfnamefont{N.}~\bibnamefont{Michel}},
  \bibinfo{author}{\bibfnamefont{W.}~\bibnamefont{Nazarewicz}},
  \bibinfo{author}{\bibfnamefont{M.}~\bibnamefont{Ploszajczak}},
  \bibnamefont{and}
  \bibinfo{author}{\bibfnamefont{J.}~\bibnamefont{Okolowicz}},
  \bibinfo{journal}{Phys. Rev. C} \textbf{\bibinfo{volume}{67}}
  (\bibinfo{year}{2003}).

\bibitem[{\citenamefont{Forss\'en et~al.}(2014)\citenamefont{Forss\'en,
  Lundmark, and Rotureau}}]{forssen2015proc}
\bibinfo{author}{\bibfnamefont{C.}~\bibnamefont{Forss\'en}},
  \bibinfo{author}{\bibfnamefont{R.}~\bibnamefont{Lundmark}}, \bibnamefont{and}
  \bibinfo{author}{\bibfnamefont{J.}~\bibnamefont{Rotureau}}
  (\bibinfo{year}{2014}), \bibinfo{note}{in Proceedings of "Dynamics of
  Critically Stable Quantum Few-Body Systems (Critical Stability 2014)", edited
  by T. Frederico (in preparation).}

\bibitem[{\citenamefont{Girardeau}(1960)}]{Girardeau:1960ff}
\bibinfo{author}{\bibfnamefont{M.}~\bibnamefont{Girardeau}},
  \bibinfo{journal}{J. Math. Phys.} \textbf{\bibinfo{volume}{1}},
  \bibinfo{pages}{516} (\bibinfo{year}{1960}).

\bibitem[{\citenamefont{Gharashi and Blume}(2013)}]{Gharashi:2013dg}
\bibinfo{author}{\bibfnamefont{S.~E.} \bibnamefont{Gharashi}} \bibnamefont{and}
  \bibinfo{author}{\bibfnamefont{D.}~\bibnamefont{Blume}},
  \bibinfo{journal}{Phys. Rev. Lett.} \textbf{\bibinfo{volume}{111}},
  \bibinfo{pages}{045302} (\bibinfo{year}{2013}).

\bibitem[{\citenamefont{Lindgren et~al.}(2014)\citenamefont{Lindgren, Rotureau,
  Forss{\'e}n, Volosniev, and Zinner}}]{Lindgren:2014kl}
\bibinfo{author}{\bibfnamefont{E.~J.} \bibnamefont{Lindgren}},
  \bibinfo{author}{\bibfnamefont{J.}~\bibnamefont{Rotureau}},
  \bibinfo{author}{\bibfnamefont{C.}~\bibnamefont{Forss{\'e}n}},
  \bibinfo{author}{\bibfnamefont{A.~G.} \bibnamefont{Volosniev}},
  \bibnamefont{and} \bibinfo{author}{\bibfnamefont{N.~T.}
  \bibnamefont{Zinner}}, \bibinfo{journal}{New J. Phys.}
  \textbf{\bibinfo{volume}{16}}, \bibinfo{pages}{063003}
  (\bibinfo{year}{2014}).

\end{thebibliography}
\bibliographystyle{apsrev}

\end{document}